\pdfoutput=1

\documentclass[
  reprint,            % Two-column journal-style layout
  superscriptaddress, % Author affiliations marked by superscripts
  amsmath,            % AMS math support
  amssymb,            % AMS math symbols
  aps,                % APS journal style
  prb,                % Physical Review B style
  floatfix            % Let REVTeX resolve difficult float placement
]{revtex4-2}

\usepackage[english]{babel}
\usepackage{graphicx}      % Include figure files
\graphicspath{{Figures/}}
\usepackage{hyperref}      % Hyperlinks
\usepackage[capitalise]{cleveref}% Context-aware cross-references
\usepackage{multirow}      % Multi-row cells in tables
\usepackage{makecell}      % Useful cell formatting in tables
\usepackage{booktabs}

\begin{document}

\title{Magnetization alignment in spin-transfer-torque magnetic random-access memory}

\author{Afan Terko}
\email{ata159@sfu.ca}
\affiliation{Department of Physics, Simon Fraser University, Burnaby, British Columbia V5A 1S6, Canada}

\author{George Lertzman-Lepofsky}
\email{gmlertzm@sfu.ca}
\affiliation{Department of Physics, Simon Fraser University, Burnaby, British Columbia V5A 1S6, Canada}

\author{Dieter Suess}
\affiliation{Faculty of Physics, University of Vienna, Vienna 1010, Austria}

\author{Claas Abert}
\email{claas.abert@univie.ac.at}
\affiliation{Faculty of Physics, University of Vienna, Vienna 1010, Austria}

\author{Erol Girt}
\email{egirt@sfu.ca}
\affiliation{Department of Physics, Simon Fraser University, Burnaby, British Columbia V5A 1S6, Canada}

\begin{abstract}
Reliable operation of perpendicular spin-transfer-torque magnetic random-access memory (p-STT-MRAM) requires control of magnetic alignment within the synthetic antiferromagnet (SAF) reference layer. At nanopillar dimensions, however, devices can exhibit magnetic states that are absent in extended thin films. We present a systematic micromagnetic study of $30~\mathrm{nm}$-diameter three-layer p-STT-MRAM nanopillars using experimentally motivated material parameters, and map equilibrium states as functions of bilinear and biquadratic interlayer exchange coupling. Phase diagrams show that introducing asymmetry between the SAF layers in saturation magnetization, anisotropy, and thickness reduces the coupling strength required to stabilize antiparallel SAF states and suppress competing configurations. Minimum-energy path calculations show that, for noncollinear antiparallel SAF states, increasing SAF asymmetry can raise SAF reversal barriers while lowering the free-layer barrier; this trade-off is absent for collinear antiparallel SAF states. Stray fields also significantly modify both SAF and free-layer energy barriers. To support the design of p-STT-MRAM devices with either collinear or noncollinear antiparallel SAF reference states, we publicly release the simulation dataset covering 4374 distinct device configurations.
\end{abstract}

\maketitle

\section{Introduction}

Spin-transfer-torque magnetic random-access memory (STT-MRAM)~\cite{slonczewski_current-driven_1996,berger_emission_1996} is a leading candidate for next-generation non-volatile memory because it combines high speed, high endurance, and scalability~\cite{khvalkovskiy_basic_2013,rizzo_fully_2013,chung_4gbit_2016}. However, achieving the required balance between low switching current and high thermal stability remains a significant challenge for high-density applications~\cite{kent_new_2015,worledge_write-error-rate_2023}.

To address these challenges, modern devices employ perpendicular STT-MRAM (p-STT-MRAM) cells~\cite{carcia_perpendicular_1985}, in which the ferromagnetic layers have perpendicular magnetic anisotropy. In a typical nanopillar stack, FM1 and FM2 form a synthetic antiferromagnet (SAF) that serves as the fixed reference layer, while FM3 serves as the free layer whose reversal encodes the binary state. Within the SAF, FM1 and FM2 are antiferromagnetically coupled across a nonmagnetic spacer. This design reduces the stray field acting on the free layer and improves the stability of the reference layer~\cite{fullerton_spintronics_2016,apalkov_magnetoresistive_2016,arora_magnetic_2017}.

The SAF alignment is governed by interlayer exchange coupling (IEC)~\cite{grunberg_layered_1986,parkin_systematic_1991}. Although IEC is often modeled primarily through the bilinear coupling coefficient, $J_1$, spacer modifications can make the biquadratic coefficient, $J_2$, significant. Such modifications include alloying with dilute magnetic impurities~\cite{winther_antiferromagnetic_2024} and high-temperature annealing required for device integration~\cite{mckinnon_thermally_2022,nunn_control_2020,abert_origin_2022,nunn_controlling_2023,lisik_noncollinear_2023}. Both alloying and annealing can weaken antiferromagnetic coupling and promote noncollinear alignment by increasing the relative contribution of $J_2$~\cite{lisik_noncollinear_2023,lertzman-lepofsky_energy_2024,wadge_modeling_2024}. These effects motivate a systematic examination of both $J_1$ and $J_2$ across the coupling strengths accessible in experimentally realizable structures.

Despite the SAF's central role in setting the reference state, most p-STT-MRAM studies have focused on optimizing the free layer. The orientations of the magnetic moments in FM1 and FM2 of the SAF are typically inferred from $M(H)$ magnetometry on extended thin films~\cite{stiles_interlayer_2005,nunn_control_2020}. In such films, magnetization reversal is dominated by domain-wall nucleation and propagation rather than the coherent rotation expected in nanoscale pillars. In patterned nanopillars, stray fields from the individual magnetic layers can further alter the stable alignment. Film-based $M(H)$ measurements alone do not determine the full set of magnetic configurations that may become stable once the material is patterned to device dimensions.

This disconnect between thin-film characterization and device-scale magnetic states leaves an important knowledge gap that can affect reliability and performance optimization. To bridge this gap, we perform finite-element micromagnetic simulations of $30~\mathrm{nm}$ diameter pillars in a three-layer (FM1/FM2/FM3) system. The simulations use material parameters from experimental measurements on state-of-the-art Co/Pt-based SAFs and CoFeB free layers. Our systematic study encompasses 33.6 million simulations across 4374 distinct device configurations and a wide range of $J_1$ and $J_2$. Within experimentally accessible coupling regimes, we observe a broad spectrum of magnetic configurations, many of which have not yet been systematically characterized. These findings provide design guidance for engineering SAFs in p-STT-MRAM, and we provide a public-access website for interactive viewing and plotting of the simulation results~\cite{terko_sfu_2026}.

The remainder of this paper is organized as follows. We first describe the micromagnetic model, nanopillar geometry, and layer parameters. We also describe the procedure used to identify equilibrium states and calculate reversal barriers between those states. We then present phase diagrams of the equilibrium configurations, examine how these configurations vary across layer parameters, and analyze the FM3 and SAF reversal energy barriers. Finally, we summarize the main design-relevant trends and discuss their implications for reliable p-STT-MRAM stacks.

\section{Methods}
\label{sec:methods}
\subsection{Micromagnetic simulation framework}

Our study relies on the continuum theory of micromagnetism, in which the magnetization is described by a continuous vector field $\mathbf{M}(\mathbf{x}) = M_s \mathbf{m}(\mathbf{x})$, where $M_s$ is the saturation magnetization and $\mathbf{m}(\mathbf{x})$ is the normalized magnetization vector ($|\mathbf{m}| = 1$)~\cite{abert_micromagnetics_2019}. We employ the finite-element method (FEM) solver \texttt{magnum.pi}~\cite{abert_magnumfe_2013,abert_micromagnetics_2019,suess_accurate_2023} to determine equilibrium magnetic configurations.

The magnetization dynamics are governed by the Landau--Lifshitz--Gilbert (LLG) equation. We seek static equilibrium configurations rather than dynamical switching. Accordingly, we relax each initial state using purely dissipative (precession-free) dynamics by omitting the precessional term and integrating
\begin{equation}
\frac{\partial \mathbf{m}}{\partial t}
= -\frac{\gamma \alpha}{1+\alpha^2}\,
\mathbf{m}\times\!\left(\mathbf{m}\times \mathbf{H}_{\mathrm{eff}}\right),
\label{eq:llg_overdamped}
\end{equation}
where $\alpha$ is the Gilbert damping parameter and $\gamma$ is the reduced gyromagnetic ratio ($\gamma \approx 2.2128 \times 10^5~\mathrm{m\,A^{-1}\,s^{-1}}$).

This formulation is equivalent to steepest-descent relaxation on the micromagnetic energy landscape subject to the constraint $|\mathbf{m}|=1$. Since the precessional term is omitted, $\alpha$ only rescales the relaxation rate and does not affect the final equilibrium state; we set $\alpha=1.0$ to maximize the prefactor and accelerate convergence.

The effective field $\mathbf{H}_{\mathrm{eff}}$ is derived from the total magnetic energy $E_{\mathrm{tot}}$:
\begin{equation}
\mathbf{H}_{\mathrm{eff}} = -\frac{1}{\mu_0 M_s} \frac{\delta E_{\mathrm{tot}}}{\delta \mathbf{m}},
\end{equation}
where $M_s$ denotes the saturation magnetization of the corresponding ferromagnetic layer. This study focuses on static equilibrium configurations corresponding to local energy minima at zero temperature ($T = 0~\mathrm{K}$); thermal fluctuations are neglected.

The total magnetic energy $E_{\mathrm{tot}}$ includes contributions from Heisenberg exchange, uniaxial magnetocrystalline anisotropy, the demagnetizing field, and interlayer exchange coupling (IEC):
\begin{align}
E_{\mathrm{tot}} = \sum_{i = 1, 2, 3} \int_{V_i} \Big[ &A_{\mathrm{ex},i}(\nabla \mathbf{m}_i)^2 - K_{u,i} (\mathbf{m}_i \cdot \hat{\mathbf{z}})^2 \nonumber \\
&- \frac{\mu_0}{2} M_{s,i} \mathbf{m}_i \cdot \mathbf{H}_{\mathrm{d}} \Big] \,\mathrm{d}V \nonumber \\
&+ \int_{S_{12}} E_{\mathrm{IEC}}(\mathbf{m}_1, \mathbf{m}_2) \,\mathrm{d}S.
\label{eq:energy}
\end{align}
Here the index $i$ labels the ferromagnetic layers (FM$i$), each characterized by exchange stiffness $A_{\mathrm{ex},i}$, uniaxial anisotropy constant $K_{u,i}$ with easy axis along $\hat{\mathbf{z}}$, and saturation magnetization $M_{s,i}$. The vector $\mathbf{m}_{i}$ is the normalized magnetization in layer $i$, and $\mathbf{H}_{\mathrm{d}}$ is the demagnetizing field.

The demagnetizing field $\mathbf{H}_{\mathrm{d}}$ accounts for long-range magnetostatic interactions, including both intralayer shape anisotropy and interlayer dipolar coupling. We compute $\mathbf{H}_{\mathrm{d}}$ using a hybrid finite-element/boundary-element (FEM/BEM) method, which accurately handles the open boundary conditions of the magnetostatic problem~\cite{fredkin_hybrid_1990,abert_micromagnetics_2019}.

The interlayer exchange coupling contribution is written in terms of the areal
energy density $E_{\mathrm{IEC}}$, integrated over the FM1/FM2 interface
$S_{12}$. We model this contribution phenomenologically using bilinear ($J_1$)
and biquadratic ($J_2$) coupling terms:
\begin{equation}
E_{\mathrm{IEC}}
= J_1 (\mathbf{m}_1 \cdot \mathbf{m}_2)
+ J_2 (\mathbf{m}_1 \cdot \mathbf{m}_2)^2 .
\end{equation}
With this sign convention, positive $J_1$ favors antiparallel alignment, while
positive $J_2$ favors perpendicular alignment. Nonmagnetic spacer regions are
modeled as vacuum.

\subsection{Geometry and material parameters}

\begin{figure}[h]
    \centering
    \includegraphics[width=0.98\columnwidth]{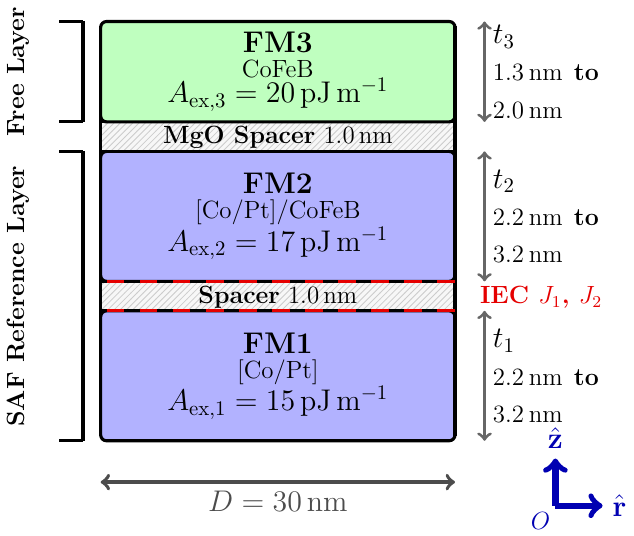}
    \caption{Stack geometry of the cylindrical nanopillar (cross-sectional view). FM1 and FM2 form the synthetic antiferromagnet (SAF) reference layer, coupled via interlayer exchange (IEC, red dashed line) with coupling constants $J_1$ and $J_2$. FM3 is the free layer. The coordinate system has $\hat{\mathbf{z}}$ perpendicular to the film plane. Layer thicknesses, saturation magnetizations, and anisotropy constants are listed in \cref{tab:combined_material_parameters}.}
    \label{fig:stack_geometry}
\end{figure}

We simulate a cylindrical nanopillar with a diameter of $30~\mathrm{nm}$ (\cref{fig:stack_geometry}). The stack consists of FM1 / IEC spacer ($1.0~\mathrm{nm}$) / FM2 / MgO barrier ($1.0~\mathrm{nm}$) / FM3. Here, FM1 is the bottom SAF layer, FM2 is the top SAF layer nearest to FM3, and FM3 is the free layer. Material parameters were selected from experimental literature on state-of-the-art p-STT-MRAM stacks. Each ferromagnetic layer is treated as an effective homogeneous magnetic layer with thickness-averaged material parameters.

The SAF layers are based on $[\mathrm{Co}/\mathrm{Pt}]_n$ multilayers, which exhibit high perpendicular magnetic anisotropy (PMA) and retain robust magnetic properties after annealing at $400\,^{\circ}\mathrm{C}$~\cite{young_lee_high_2013}. FM1 consists solely of $[\mathrm{Co}/\mathrm{Pt}]_n$, whereas FM2 comprises $[\mathrm{Co}/\mathrm{Pt}]_n$ together with a CoFeB polarization-enhancement layer adjacent to the MgO barrier. The exchange stiffness values are $A_{\mathrm{ex},1} = 15~\mathrm{pJ\,m^{-1}}$ for FM1 and $A_{\mathrm{ex},2} = 17~\mathrm{pJ\,m^{-1}}$ for FM2~\cite{devolder_exchange_2016,eyrich_effects_2014}.

For FM1 and FM2, the effective material parameters were varied independently within ranges reported for annealed $[\mathrm{Co}/\mathrm{Pt}]_n$ multilayers~\cite{lee_effects_2013,lim_effect_2015,young_lee_high_2013,yakushiji_very_2017,sato_properties_2014,tomczak_influence_2016,yakushiji_perpendicular_2015} and CoFeB~\cite{huang_electric_2016,bersweiler_magnetic_2017,iwata-harms_ultrathin_2019}. The independently varied parameters were the saturation magnetization $M_s = 900, 1000, 1100~\mathrm{kA\,m^{-1}}$, the uniaxial anisotropy constant $K_u = 0.80, 0.95, 1.10~\mathrm{MJ\,m^{-3}}$, and the layer thickness $t = 2.2, 2.7, 3.2~\mathrm{nm}$.

The free layer, FM3, is based on a dual-MgO-interface CoFeB structure~\cite{huang_electric_2016,bersweiler_magnetic_2017,iwata-harms_ultrathin_2019}. We fixed the exchange stiffness at $A_{\mathrm{ex},3} = 20~\mathrm{pJ\,m^{-1}}$~\cite{devolder_spin-torque_2011,sato_perpendicular-anisotropy_2012} and varied the saturation magnetization $M_s = 1000, 1300, 1600~\mathrm{kA\,m^{-1}}$ and thickness $t = 1.3, 2.0~\mathrm{nm}$~\cite{santos_ultrathin_2020,couet_impact_2017,choi_measurement_2023,lee_effect_2016,iwata-harms_high-temperature_2018,lang_effect_2025}.

Thermal stability factors were estimated using $\Delta = K_{\mathrm{eff}}V/(k_{\mathrm{B}}T)$ at $T = 300~\mathrm{K}$, where $V$ is the magnetic layer volume. The effective anisotropy is written as $K_{\mathrm{eff}} = K_u + K_{\mathrm{shape}}$, where the shape-anisotropy contribution is
\begin{align}
K_{\mathrm{shape}} = \frac{1}{2}\mu_0 M_s^2 (N_{\perp} - N_z).
\label{eq:shape}
\end{align}
With this convention, $K_{\mathrm{shape}}<0$ for thin perpendicular cylinders, so shape anisotropy reduces the effective perpendicular anisotropy. Here, $N_z$ is the demagnetization factor along $\hat{\mathbf{z}}$ and $N_{\perp}$ is the in-plane demagnetization factor. For the circular cylinders considered here, $N_x = N_y = N_{\perp}$. The magnetometric demagnetization factors for finite cylinders were calculated using the expressions given by Joseph~\cite{joseph_ballistic_1966}.

For FM1 and FM2, the resulting macrospin stability factors span $\Delta_1, \Delta_2 = 85$--$411$, reflecting the high stability required of the SAF reference layer. For FM3, $K_u$ was chosen separately for each $(M_s, t)$ combination to maintain a fixed free-layer stability factor $\Delta_3 = 60$. This choice is motivated by the Arrhenius--Néel thermal activation model, in which the retention time is $\tau = \tau_0 \exp(K_{\mathrm{eff}}V/k_{\mathrm{B}}T) = \tau_0 \exp(\Delta)$. Taking $\tau_0 = 1~\mathrm{ns}$ gives $\Delta = \ln(\tau/\tau_0) \approx 40$ for 10-year single-bit retention; however, high-density memory arrays require larger stability factors, of order $\Delta \ge 60$, to suppress data loss~\cite{kent_new_2015,809134}. These values define the nominal macrospin thermal stability of the individual layers. In the full nanopillar, magnetostatic interactions between layers can shift the effective reversal barriers, which are evaluated separately using the minimum-energy path method described in \cref{sec:mep_method}. The representative parameter sets are summarized in \cref{tab:combined_material_parameters}.

\begin{table}[htbp]
\centering
\caption{Representative material parameters and thermal stability factors for SAF layers (FM1/FM2) and the free layer (FM3). The effective anisotropy $K_{\mathrm{eff}} = K_u + K_{\mathrm{shape}}$ for FM1/FM2 is computed using the shape-anisotropy contribution in \cref{eq:shape}, with demagnetization factors for a $30~\mathrm{nm}$ diameter cylinder~\cite{joseph_ballistic_1966}. For FM3, the uniaxial anisotropy $K_u$ is selected to achieve a fixed thermal stability $\Delta = 60$. The thermal stability factor is evaluated at $T = 300~\mathrm{K}$.}
\label{tab:combined_material_parameters}
\setlength{\tabcolsep}{6pt}
\renewcommand{\arraystretch}{1.05}
\begin{tabular}{c|ccccc}
\toprule
Layer & $\Delta$ & \makecell{$M_s$ \\ ($\mathrm{kA\,m^{-1}}$)} & \makecell{$K_u$ \\ ($\mathrm{MJ\,m^{-3}}$)} & \makecell{$t$ \\ ($\mathrm{nm}$)} & \makecell{$K_{\mathrm{eff}}$ \\ ($\mathrm{MJ\,m^{-3}}$)} \\
\midrule
\multirow{8}{*}{\rotatebox[origin=c]{90}{FM1/FM2}}
 & 85   & 1100 & 0.8 & 2.2 & 0.226 \\
 & 154  & 1100 & 0.8 & 3.2 & 0.282 \\
 & 156  & 900  & 0.8 & 2.2 & 0.416 \\
 & 198  & 1100 & 1.1 & 2.2 & 0.526 \\
 & 248  & 900  & 0.8 & 3.2 & 0.453 \\
 & 269  & 900  & 1.1 & 2.2 & 0.716 \\
 & 318  & 1100 & 1.1 & 3.2 & 0.582 \\
 & 411  & 900  & 1.1 & 3.2 & 0.753 \\
\midrule
\multirow{2}{*}{\rotatebox[origin=c]{90}{FM3}}
 & 60 & 1000 & 0.794 & 1.3 & 0.270 \\
 & 60 & 1600 & 1.416 & 2.0 & 0.176 \\
\bottomrule
\end{tabular}
\end{table}

\subsection{Micromagnetic simulation setup}

All FEM mesh sizes were kept below the smallest exchange length in our parameter space, $\ell_{\mathrm{ex}} = \sqrt{A_{\mathrm{ex}}/K_{\mathrm{eff}}}$~\cite{abert_micromagnetics_2019,hubert_magnetic_1998}. In the film plane, we used a mesh size of $3~\mathrm{nm}$. Along $\hat{\mathbf{z}}$, the mesh size was set to the layer thickness $t$. We verified that finer meshes produced the same equilibrium configurations. Each simulation was relaxed to a stationary state using \cref{eq:llg_overdamped}. Convergence was confirmed by the absence of further changes in the total energy and layer-averaged magnetization angles.

We investigated $27 \times 27 \times 6 = 4374$ distinct device configurations ($3\,M_s \times 3\,K_u \times 3\,t = 27$ parameter sets each for FM1 and FM2, and $3\,M_s \times 2\,t = 6$ for FM3). For each configuration, $J_1$ and $J_2$ were varied from 0 to $3.0~\mathrm{mJ\,m^{-2}}$ in increments of $0.1~\mathrm{mJ\,m^{-2}}$, giving a $31 \times 31$ grid (961 coupling combinations). Interactive visualizations for all configurations are available online~\cite{terko_sfu_2026}. We discuss only a representative subset of $8 \times 8 \times 2 = 128$ configurations, as listed in \cref{tab:combined_material_parameters}.

To search for equilibrium magnetic configurations, the magnetization of each layer was initialized along $\pm\hat{\mathbf{z}}$ in eight combinations, where the arrow sequence lists the initial magnetization directions of FM1, FM2, and FM3:
\[
(\uparrow\uparrow\uparrow),\; (\uparrow\uparrow\downarrow),\; (\uparrow\downarrow\uparrow),\; (\uparrow\downarrow\downarrow),\; (\downarrow\uparrow\uparrow),\; (\downarrow\uparrow\downarrow),\; (\downarrow\downarrow\uparrow),\; (\downarrow\downarrow\downarrow).
\]
A small polar angle $\theta = 0.1^{\circ}$ from $\hat{\mathbf{z}}$ was applied to each layer to break symmetry. Each initial state was then relaxed to obtain an equilibrium configuration, which was classified according to \cref{subsec:class}. In total, this yields $4374 \times 961 \times 8 \approx 33.6$ million micromagnetic simulations.

To verify that the eight perpendicular initial states capture all energy minima, we repeated the relaxation procedure for a representative subset of structures and $(J_1, J_2)$ pairs using 1000 random initial configurations. In each case, every layer's magnetization was assigned randomly chosen polar and azimuthal angles, $\theta$ and $\phi$. These runs yielded the same set of minima as the eight perpendicular states.

\subsection{Classification of magnetic configurations}
\label{subsec:class}

\begin{figure}[h]
    \centering
    \includegraphics[width=0.98\columnwidth]{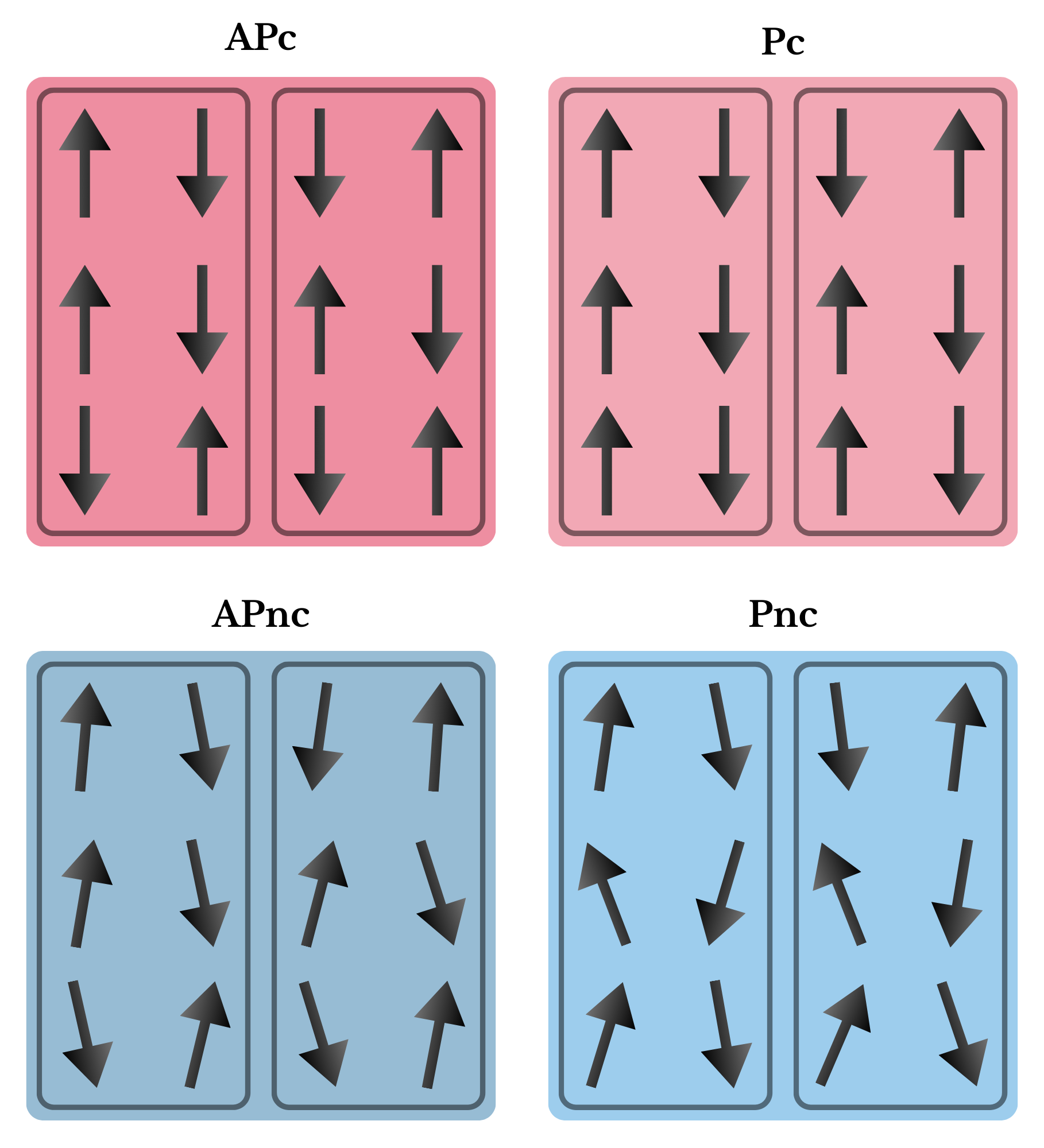}
    \caption{The 16 possible magnetic configurations of the three-layer structure (layer order FM1/FM2/FM3 from bottom to top), categorized into four configuration groups: collinear antiparallel (APc), collinear parallel (Pc), noncollinear antiparallel (APnc), and noncollinear parallel (Pnc). Within each group, states of equal energy are bounded by a thin solid line, resulting in only two distinct energy levels per group. Pnc states are also characterized by an antiparallel alignment of their in-plane magnetization components ($\Delta\phi = 180^{\circ}$), where $\Delta\phi$ is the relative azimuthal angle between the in-plane components of FM1 and FM2.}
    \label{fig:16_states}
\end{figure}

We classified relaxed equilibrium states based on the layer-averaged polar angles $\theta_1$, $\theta_2$, and $\theta_3$ of FM1, FM2, and FM3, respectively, measured from $\hat{\mathbf{z}}$. We label the SAF orientation as parallel (P) when the FM1 and FM2 magnetization components along $\hat{\mathbf{z}}$ have the same sign, $\mathrm{sgn}(\cos\theta_1) = \mathrm{sgn}(\cos\theta_2)$, and antiparallel (AP) when they have opposite signs. We further label a configuration as collinear (c) if $\theta_1$, $\theta_2$, and $\theta_3$ each fall within one degree of $0^{\circ}$ or $180^{\circ}$, and noncollinear (nc) otherwise. This yields four configuration groups: APc, APnc, Pc, and Pnc, illustrated in \cref{fig:16_states}.

The total energy in \cref{eq:energy} is invariant under reversal of all magnetic moments, i.e., under the transformation $\mathbf{m}_i \to -\mathbf{m}_i$ for all layers ($i = 1, 2, 3$). Consequently, each equilibrium state has an energetically equivalent spin-flipped counterpart (e.g., $E(\uparrow\downarrow\uparrow) = E(\downarrow\uparrow\downarrow)$). Although the classification contains 16 possible magnetic configurations, global spin-reversal symmetry pairs them into 8 energetically equivalent pairs. Depending on the specific material parameters and coupling strengths ($J_1$ and $J_2$), our simulations yield sets of 2, 4, 6, or 8 magnetic configurations, corresponding respectively to 1, 2, 3, or 4 distinct energy levels.

\subsection{Minimum-energy path calculation}
\label{sec:mep_method}

To quantify the stability of magnetic states and compare the likelihood of transitions between them, we computed the energy barriers separating local minima in the magnetic energy landscape. We used the simplified string method~\cite{e_simplified_2007}, a numerical technique that determines the minimum-energy path (MEP) connecting two stable magnetization configurations.

The calculation begins by defining an arbitrary path between the start and end configurations, discretized into a chain of $N_{\mathrm{img}}$ intermediate magnetic states (images). We used $N_{\mathrm{img}} = 100$ images and verified convergence on a representative subset by repeating calculations with a larger number of images, which produced the same barriers.

To prevent images from clustering near the minima, we applied a reparameterization step after each relaxation step. The images are redistributed along the current path using spline interpolation, ensuring equidistant spacing~\cite{e_simplified_2007}. This two-step procedure (relaxation followed by reparameterization) is repeated until the path converges. The resulting string represents the MEP, and the highest-energy image along this path approximates the saddle point. The energy barrier for the transition from state $i$ to state $j$ is defined as
\begin{equation}
    E_{\mathrm{b}}^{i \to j} = E_{\mathrm{saddle}} - E_i, \qquad E_{\mathrm{b}}^{j \to i} = E_{\mathrm{saddle}} - E_j,
    \label{eq:barrier_def}
\end{equation}
where $E_{\mathrm{saddle}}$ is the maximum energy along the MEP and $E_i$, $E_j$ are the energies of the initial and final minima, respectively. Because the two minima connected by a given MEP can have different energies, the forward and reverse barriers generally differ. For $N$ minima, there are $N(N-1)/2$ unique unordered pairs: six for $N=4$, 15 for $N=6$, and 28 for $N=8$.

\section{Results}

\subsection{Magnetic configuration phase diagrams}

\begin{figure*}[t]
    \centering
        \includegraphics[width=0.98\textwidth]{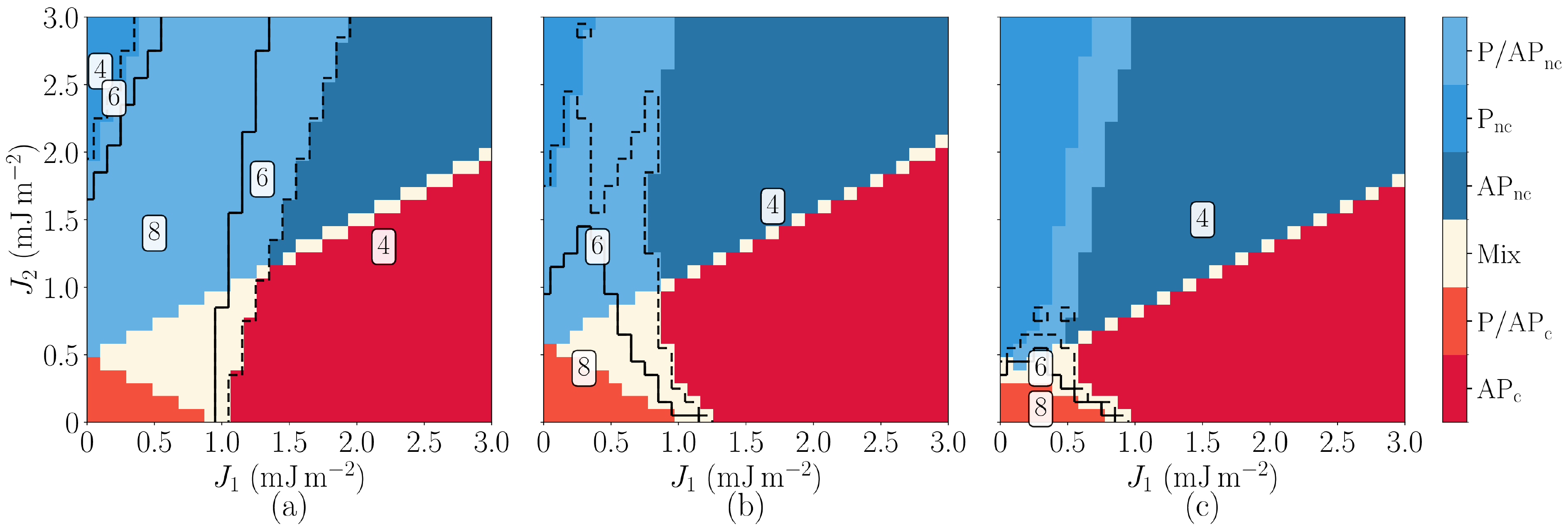}
    \caption{Magnetic-configuration phase diagrams of a three-layer structure as functions of the bilinear ($J_1$) and biquadratic ($J_2$) interlayer exchange coupling between FM1 and FM2. \textbf{(a)} $M_{s,1} = M_{s,2}$, $K_{u,1} = K_{u,2}$, $t_1 = t_2$, $\Delta_1 = \Delta_2 = 156$; \textbf{(b)} $M_{s,1} = M_{s,2}$, $K_{u,1} \neq K_{u,2}$, $t_1 = t_2$, $\Delta_1 = 269$, $\Delta_2 = 156$; \textbf{(c)} $M_{s,1} \neq M_{s,2}$, $K_{u,1} \neq K_{u,2}$, $t_1 \neq t_2$, $\Delta_1 = 85$, $\Delta_2 = 411$. Parameters of FM3 are fixed: $M_{s,3} = 1000~\mathrm{kA\,m^{-1}}$, $K_{u,3} = 0.794~\mathrm{MJ\,m^{-3}}$, and $t_3 = 1.3~\mathrm{nm}$. Red regions indicate collinear states (Pc, APc); blue regions indicate noncollinear states (Pnc, APnc); off-white denotes coexistence of collinear and noncollinear minima. Contour lines denote the number of stable magnetic configurations: $4$ outside the dashed contour, $6$ between the dashed and solid contours, and $8$ inside the solid contour. Increasing SAF asymmetry shifts the antiparallel-only boundaries toward smaller $J_1$ (and smaller $J_2$ for APnc).}
    \label{fig:magnetic_config}
\end{figure*}

We performed micromagnetic simulations to map the equilibrium magnetization configurations of the FM1/FM2/FM3 structure as functions of the bilinear ($J_1$) and biquadratic ($J_2$) interlayer exchange couplings between FM1 and FM2. In total, 4374 distinct three-layer magnetic structures were simulated; all corresponding phase diagrams are available online~\cite{terko_sfu_2026}. \Cref{fig:magnetic_config} presents phase diagrams for three representative structures that capture the trends observed across all simulations. In these simulations, the parameters of FM1 and FM2 are varied, while those of FM3 are held fixed at $M_{s,3} = 1000~\mathrm{kA\,m^{-1}}$, $K_{u,3} = 0.794~\mathrm{MJ\,m^{-3}}$, and $t_3 = 1.3~\mathrm{nm}$.

\Cref{fig:magnetic_config}(a) shows the case in which FM1 and FM2 are identical: $M_{s,1} = M_{s,2} = 900~\mathrm{kA\,m^{-1}}$, $K_{u,1} = K_{u,2} = 0.8~\mathrm{MJ\,m^{-3}}$, and $t_1 = t_2 = 2.2~\mathrm{nm}$ ($\Delta_1 = \Delta_2 = 156$). An APc-only region requires $J_1 \gtrsim 1.1~\mathrm{mJ\,m^{-2}}$, while an APnc-only region emerges only when both $J_1 \gtrsim 1.5~\mathrm{mJ\,m^{-2}}$ and $J_2 \gtrsim 1.2~\mathrm{mJ\,m^{-2}}$. At smaller $J_1$, the phase diagram exhibits mixed regions containing APc, APnc, Pc, and Pnc configurations, and for small $J_1$ and large $J_2$ only Pnc configurations are stable.

\Cref{fig:magnetic_config}(b) illustrates the effect of introducing anisotropy asymmetry while keeping the magnetization and thickness identical between FM1 and FM2 ($M_{s,1} = M_{s,2} = 900~\mathrm{kA\,m^{-1}}$; $t_1 = t_2 = 2.2~\mathrm{nm}$; $K_{u,1} = 1.1~\mathrm{MJ\,m^{-3}}$; $K_{u,2} = 0.8~\mathrm{MJ\,m^{-3}}$; $\Delta_1 = 269$, $\Delta_2 = 156$). The coupling thresholds for APc-only and APnc-only regions decrease: APc now requires $J_1 \gtrsim 0.9~\mathrm{mJ\,m^{-2}}$, and APnc requires $J_1 \gtrsim 0.9~\mathrm{mJ\,m^{-2}}$ with $J_2 \gtrsim 1.0~\mathrm{mJ\,m^{-2}}$. The mixed-configuration region shrinks accordingly, and the Pnc region expands for small $J_1$ and large $J_2$.

\Cref{fig:magnetic_config}(c) shows a structure in which FM1 and FM2 differ in all three parameters: magnetization ($M_{s,1} = 1100~\mathrm{kA\,m^{-1}}$; $M_{s,2} = 900~\mathrm{kA\,m^{-1}}$), thickness ($t_1 = 2.2~\mathrm{nm}$; $t_2 = 3.2~\mathrm{nm}$), and anisotropy ($K_{u,1} = 0.8~\mathrm{MJ\,m^{-3}}$; $K_{u,2} = 1.1~\mathrm{MJ\,m^{-3}}$), yielding $\Delta_1 = 85$ and $\Delta_2 = 411$. This fully asymmetric case substantially broadens both the APc and APnc regions: APc requires only $J_1 \gtrsim 0.65~\mathrm{mJ\,m^{-2}}$, and APnc requires $J_1 \gtrsim 0.65~\mathrm{mJ\,m^{-2}}$ with $J_2 \gtrsim 0.7~\mathrm{mJ\,m^{-2}}$. The mixed-configuration region is further reduced, with most of it containing only four magnetic configurations. The Pnc region also expands and appears for $J_1 \lesssim 0.5~\mathrm{mJ\,m^{-2}}$ and $J_2 \gtrsim 0.5~\mathrm{mJ\,m^{-2}}$.

APc-only or APnc-only coupling regions are desirable for reliable p-STT-MRAM operation because the stray fields from FM1 and FM2 acting on FM3 partially or fully cancel. Regions that support multiple configuration groups, by contrast, should be avoided. In such regions, the magnetization in p-STT-MRAM nanopillars may relax into different configuration groups (\cref{fig:16_states}), leading to variability in the magnetic state that encodes the binary information. This variability can make readout less reliable and produces inconsistent stray fields on FM3.

Results in \Cref{fig:magnetic_config} demonstrate that introducing asymmetry between FM1 and FM2 lowers the interlayer exchange coupling required to obtain APc-only or APnc-only configurations.

\subsection{Prevalence of antiparallel configurations}

The phase diagrams in \cref{fig:magnetic_config} illustrate trends for three representative structures. To summarize results across all FM1 and FM2 parameter combinations in \cref{tab:combined_material_parameters}, we quantified how the thermal stability factors $(\Delta_1, \Delta_2)$ of FM1 and FM2 affect the prevalence of APc-only or APnc-only regions.

\begin{figure*}[t]
\centering
\includegraphics[width=\textwidth]{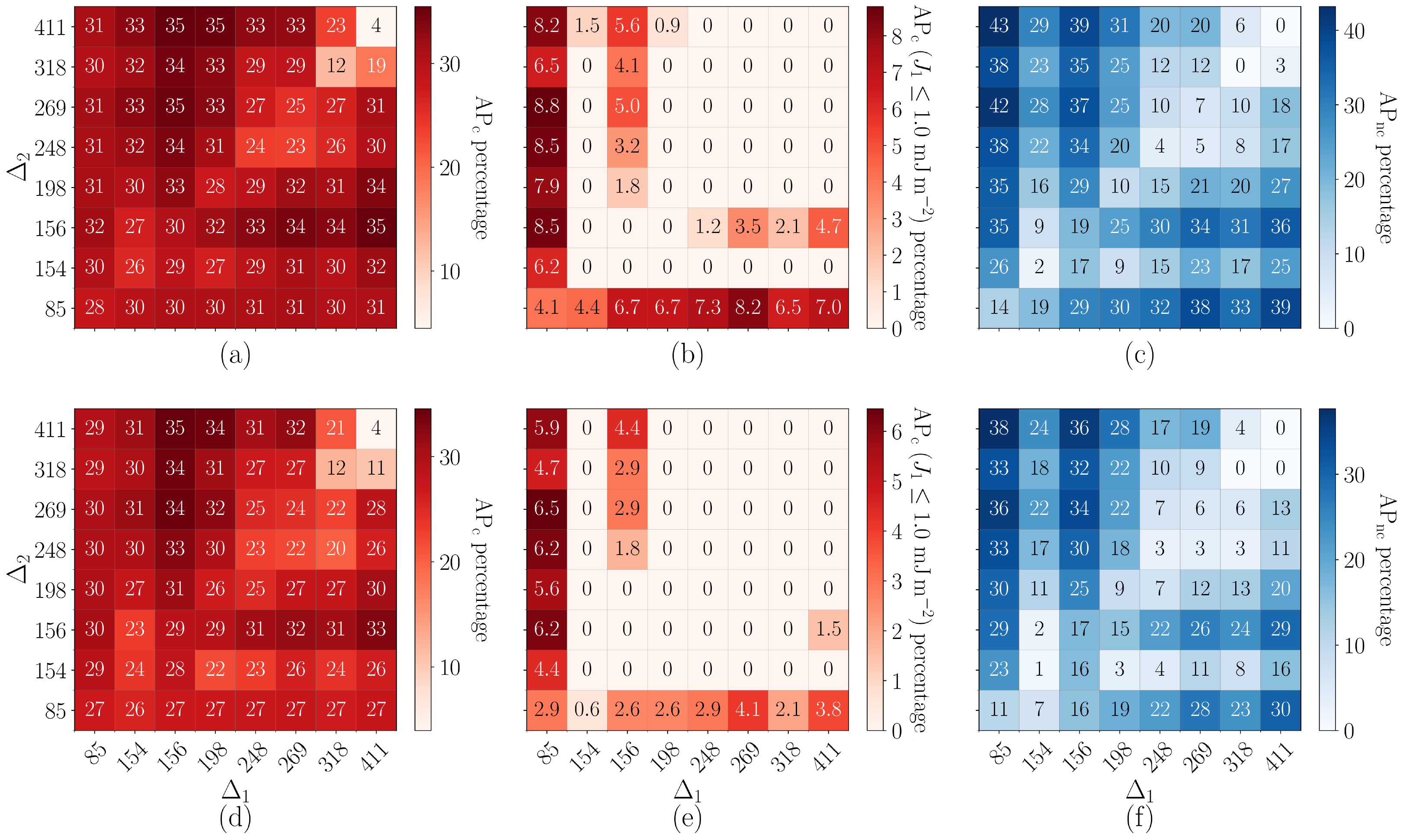}
\caption{Percentage of $(J_1, J_2)$ points in the $31\times 31$ coupling grid that yield only antiparallel magnetic configurations: APc \textbf{(a, b, d, e)} or APnc \textbf{(c, f)}. A point is counted if the relaxation from the eight seeded initial states yields exactly four stable minima, all belonging to APc or all belonging to APnc. FM1 and FM2 are chosen from the eight SAF parameter sets listed in \cref{tab:combined_material_parameters}, forming an $8\times 8$ subset of the full $27\times 27$ SAF parameter space; the complete set of phase diagrams is available online~\cite{terko_sfu_2026}. Panels \textbf{(a, c, d, f)}: $J_1, J_2 = 0$--$3.0~\mathrm{mJ\,m^{-2}}$; panels \textbf{(b, e)}: $J_1 = 0$--$1.0~\mathrm{mJ\,m^{-2}}$, $J_2 = 0$--$3.0~\mathrm{mJ\,m^{-2}}$. For \textbf{(a--c)}, FM3 parameters are $M_{s,3} = 1000~\mathrm{kA\,m^{-1}}$, $K_{u,3} = 0.794~\mathrm{MJ\,m^{-3}}$, and $t_3 = 1.3~\mathrm{nm}$. For \textbf{(d--f)}, FM3 parameters are $M_{s,3} = 1600~\mathrm{kA\,m^{-1}}$, $K_{u,3} = 1.416~\mathrm{MJ\,m^{-3}}$, and $t_3 = 2.0~\mathrm{nm}$.}
\label{fig:antiparallel_distribution}
\end{figure*}

In \cref{fig:antiparallel_distribution}, we restrict FM1 and FM2 to the eight SAF parameter sets listed in \cref{tab:combined_material_parameters}, forming an $8\times 8$ subset of the full $27\times 27$ SAF parameter space. Phase diagrams for the complete $27 \times 27$ parameter space are available online~\cite{terko_sfu_2026}. For each $(\Delta_1, \Delta_2)$ combination in this subset, we computed the percentage of $(J_1, J_2)$ points in the parameter sweep that yield exactly four stable minima, all belonging to APc or all belonging to APnc. \Cref{fig:antiparallel_distribution} summarizes these results.

In \cref{fig:antiparallel_distribution}(a, d), the APc percentage is shown over the full coupling range $(0 \leq J_1, J_2 \leq 3~\mathrm{mJ\,m^{-2}})$. The APc percentage is highest when $\Delta_1$ and $\Delta_2$ differ substantially and is lowest when both stability factors are large ($\Delta_1, \Delta_2 \geq 318$). Outside these limiting cases, the APc percentage depends only weakly on the specific values of $\Delta_1$ and $\Delta_2$.

Restricting the analysis to weak bilinear coupling $(J_1 \leq 1.0~\mathrm{mJ\,m^{-2}})$, the regime most relevant after high-temperature annealing, reveals a stricter requirement for APc configurations, as shown in \cref{fig:antiparallel_distribution}(b, e). In this regime, at least one of $\Delta_1$ or $\Delta_2$ must be small ($\lesssim 156$) for any APc-only regions to appear; when both exceed this threshold, the APc percentage drops to essentially zero.

The APnc configurations exhibit a different trend, as shown in \cref{fig:antiparallel_distribution}(c, f). The APnc fraction increases when $\Delta_1$ and $\Delta_2$ are asymmetric. For example, when one stability factor is $85$ and the other exceeds $154$, APnc-only regions occur for 19--43\% of the $(J_1, J_2)$ combinations. In contrast, when $\Delta_1$ and $\Delta_2$ are equal, APnc-only regions occur for fewer than 19\% of the $(J_1, J_2)$ combinations. A slight increase in the APnc percentage is observed when $\Delta_2 > \Delta_1$, although the opposite ordering produces similar results.

Comparing \cref{fig:antiparallel_distribution}(a--c) and \cref{fig:antiparallel_distribution}(d--f) reveals the influence of the free layer (FM3). The panels in \cref{fig:antiparallel_distribution}(d--f) correspond to a thicker, higher-$M_s$ free layer $(M_{s,3} = 1600~\mathrm{kA\,m^{-1}}$, $t_3 = 2.0~\mathrm{nm})$, whereas \cref{fig:antiparallel_distribution}(a--c) uses $M_{s,3} = 1000~\mathrm{kA\,m^{-1}}$ and $t_3 = 1.3~\mathrm{nm}$; in both cases, $K_{u,3}$ is adjusted to maintain $\Delta_3 = 60$ (\cref{tab:combined_material_parameters}). The thicker, higher-$M_s$ free layer generates a stronger stray field on the SAF, which reduces the overall percentage of APc and APnc configurations. This reduction is more pronounced for APnc than for APc, indicating that noncollinear SAF states are more sensitive to the free-layer stray field.

Taken together, these results show that the prevalence of APc and APnc configurations is governed primarily by the asymmetry between $\Delta_1$ and $\Delta_2$, with the free-layer stray field acting as a secondary factor that suppresses both APc and APnc configurations.

\subsection{Stability of FM3 and the SAF}
\label{sec:coupling_barriers}

The previous subsections identify which magnetic configurations are equilibrium states for given material parameters and interlayer couplings $(J_1, J_2)$, and determine the conditions under which APc-only or APnc-only configurations are stable. Reliable p-STT-MRAM operation also requires that the free-layer reversal barrier be high enough for data retention. We use $\Delta_3 = 60$ as the isolated-FM3 design target, corresponding to $E_{\mathrm{b}} = 60\,k_{\mathrm{B}}T$ at $T = 300~\mathrm{K}$. In the three-layer stack, only FM1 and FM2 are exchange-coupled via $(J_1, J_2)$, while FM3 is dipolar-coupled to the SAF; these dipolar fields can shift the actual FM3 barrier away from this target. Moreover, the SAF reversal barrier should remain larger than that of FM3 to suppress SAF reversal during FM3 switching. In this subsection, we evaluate these barriers as a function of $(J_1, J_2)$ for three sets of SAF layer parameters.

\begin{figure*}[t]
    \centering
    \includegraphics[width=0.98\textwidth]{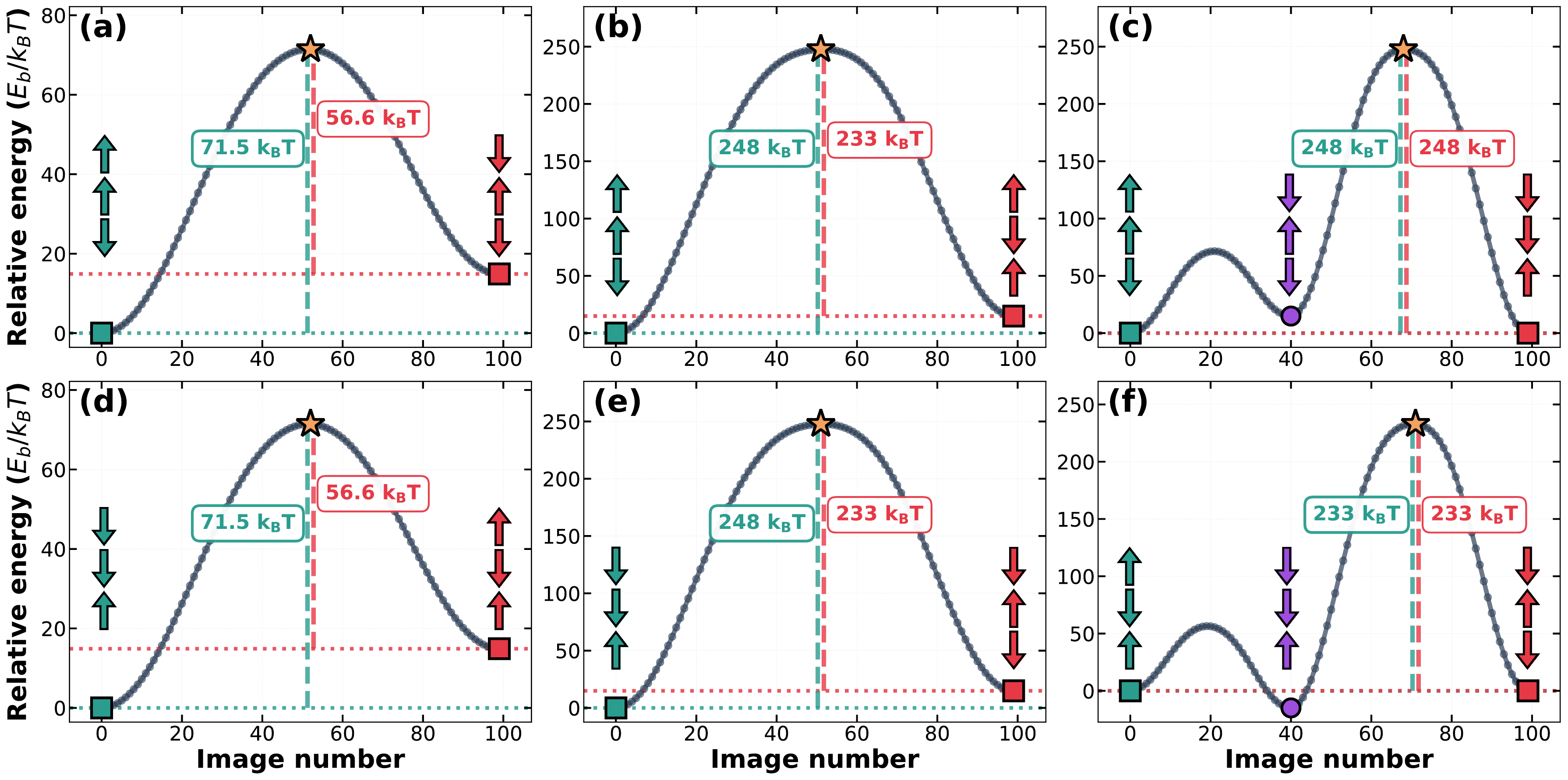}
    \caption{Energy profiles along minimum-energy paths (MEPs) for all transitions between the four APc minima, calculated via the string method with 100 images. Parameters: $\Delta_1 = \Delta_2 = 156$, $\Delta_3 = 60$, $M_{s,3} = 1000~\mathrm{kA\,m^{-1}}$, $t_3 = 1.3~\mathrm{nm}$, $J_1 = 1.5~\mathrm{mJ\,m^{-2}}$, $J_2 = 0.5~\mathrm{mJ\,m^{-2}}$. Energies are computed from the micromagnetic energy without thermal fluctuations and normalized by $k_{\mathrm{B}}T$ at $T = 300~\mathrm{K}$. \textbf{(a, d)} FM3 reversal, \textbf{(b, e)} SAF reversal, \textbf{(c, f)} reversal of all three layers. Green and red arrows indicate the initial and final magnetic configurations; violet arrows mark intermediate configurations in panels~\textbf{(c, f)}. Yellow stars mark the saddle points.}
    \label{fig:barriers}
\end{figure*}

We compute energy barriers along minimum-energy paths (MEPs) using the string method (\cref{sec:mep_method}) and report them in units of $k_{\mathrm{B}}T$ at $T = 300~\mathrm{K}$ for direct comparison with the single-layer thermal stability factors $\Delta_i$ in \cref{tab:combined_material_parameters}. As a starting point, we consider a symmetric SAF ($\Delta_1 = \Delta_2 = 156$) and FM3 with $M_{s,3} = 1000~\mathrm{kA\,m^{-1}}$ and $t_3 = 1.3~\mathrm{nm}$ ($\Delta_3 = 60$). The interlayer couplings are set to $J_1 = 1.5~\mathrm{mJ\,m^{-2}}$ and $J_2 = 0.5~\mathrm{mJ\,m^{-2}}$, which places the system in the APc-only regime of \cref{fig:magnetic_config}(a), yielding four APc minima. The MEPs and corresponding barriers for all transitions between these four minima are shown in \cref{fig:barriers}.

The four APc minima, written in FM1/FM2/FM3 order, are $(\uparrow\downarrow\uparrow)$, $(\uparrow\downarrow\downarrow)$, $(\downarrow\uparrow\uparrow)$, and $(\downarrow\uparrow\downarrow)$ (\cref{fig:16_states}). These four minima yield six transition pairs, each with two barrier heights, $E_{\mathrm{b}}^{i \to j}$ and $E_{\mathrm{b}}^{j \to i}$, defined in \cref{eq:barrier_def}. \Cref{fig:barriers} groups the transitions by which layers reverse. Panels~(a, d) show FM3 reversal: $(\downarrow\uparrow\uparrow) \leftrightarrow (\downarrow\uparrow\downarrow)$ and $(\uparrow\downarrow\downarrow) \leftrightarrow (\uparrow\downarrow\uparrow)$. Panels~(b, e) show SAF reversal: $(\downarrow\uparrow\uparrow) \leftrightarrow (\uparrow\downarrow\uparrow)$ and $(\uparrow\downarrow\downarrow) \leftrightarrow (\downarrow\uparrow\downarrow)$. Panels~(c, f) show transitions in which all three layers reverse: $(\downarrow\uparrow\uparrow) \leftrightarrow (\uparrow\downarrow\downarrow)$ and $(\uparrow\downarrow\uparrow) \leftrightarrow (\downarrow\uparrow\downarrow)$. The energy is invariant under simultaneous reversal of all layer magnetizations (\cref{subsec:class}), so panels~(a) and~(d) share the same MEP and energy profile, as do panels~(b) and~(e). The MEPs in panels~(c, f) pass through intermediate minima, because the FM3 reversal barrier is much lower than that of the SAF, and therefore contain two saddle points. Consistent with \cref{eq:barrier_def}, the reported barrier for each full path is referenced to the maximum-energy point along that MEP.

\Cref{fig:barriers}(a, d) shows the FM3 reversal barrier. As expected, it is around $60\,k_{\mathrm{B}}T$, consistent with the value for an isolated FM3. However, the SAF stray field on FM3 does not fully cancel and can either oppose or assist FM3 reversal, so that $E_{\mathrm{b}}^{\downarrow\uparrow\uparrow \to \downarrow\uparrow\downarrow} > 60\,k_{\mathrm{B}}T > E_{\mathrm{b}}^{\downarrow\uparrow\downarrow \to \downarrow\uparrow\uparrow}$. For the symmetric SAF considered here, FM1 and FM2 have identical thickness and saturation magnetization, but FM2 is closer to FM3. The dipolar field decays with separation, so FM2 dominates the net SAF field at FM3 and the more distant FM1 only partially cancels its contribution. In the transition $(\downarrow\uparrow\uparrow) \to (\downarrow\uparrow\downarrow)$, the net SAF field opposes FM3 reversal, yielding $E_{\mathrm{b}}^{\downarrow\uparrow\uparrow \to \downarrow\uparrow\downarrow} = 71.5\,k_{\mathrm{B}}T$. In the reverse transition $(\downarrow\uparrow\downarrow) \to (\downarrow\uparrow\uparrow)$, the net SAF field assists reversal, reducing the barrier to $E_{\mathrm{b}}^{\downarrow\uparrow\downarrow \to \downarrow\uparrow\uparrow} = 56.6\,k_{\mathrm{B}}T$.

\Cref{fig:barriers}(b, e) shows the SAF reversal barrier, i.e., the barrier for reversing both FM1 and FM2 while the magnetization of FM3 remains unchanged. In addition to $\Delta_1$ and $\Delta_2$, the antiferromagnetic interlayer exchange coupling between FM1 and FM2 raises the SAF reversal barrier~\cite{10.1063/1.1467977}. For the parameters considered here, the SAF barrier is more than three times higher than the FM3 reversal barrier. The stray field from FM3 also shifts the SAF barrier: it opposes SAF reversal for $(\downarrow\uparrow\uparrow) \to (\uparrow\downarrow\uparrow)$, yielding $E_{\mathrm{b}}^{\downarrow\uparrow\uparrow \to \uparrow\downarrow\uparrow} = 248\,k_{\mathrm{B}}T$, and it assists SAF reversal for $(\downarrow\uparrow\downarrow) \to (\uparrow\downarrow\downarrow)$, giving $E_{\mathrm{b}}^{\downarrow\uparrow\downarrow \to \uparrow\downarrow\downarrow} = 233\,k_{\mathrm{B}}T$.

\Cref{fig:barriers}(c, f) shows the two remaining transitions, in which all three layers reverse. In both panels, FM3 switches first and the SAF follows, producing two saddle points along each MEP, because the FM3 barrier is much lower than the SAF barrier for the chosen layer parameters and $(J_1, J_2)$. The barrier heights in the two panels differ, reflecting the stray-field effects on both FM3 and the SAF discussed above. Together, the six MEPs show that, for $(\Delta_1, \Delta_2) = (156, 156)$ and $(J_1, J_2) = (1.5, 0.5)~\mathrm{mJ\,m^{-2}}$, the SAF barrier is much larger than that of FM3. They also show that stray fields from the SAF modify the FM3 reversal barriers, while stray fields from FM3 modify the SAF reversal barriers, depending on the relative magnetization alignment between the SAF and FM3. We next examine how these barriers evolve across the full $(J_1, J_2)$ space.

\subsubsection*{Switching-barrier maps}
\begin{figure*}[t]
    \centering
    \includegraphics[width=0.99\textwidth]{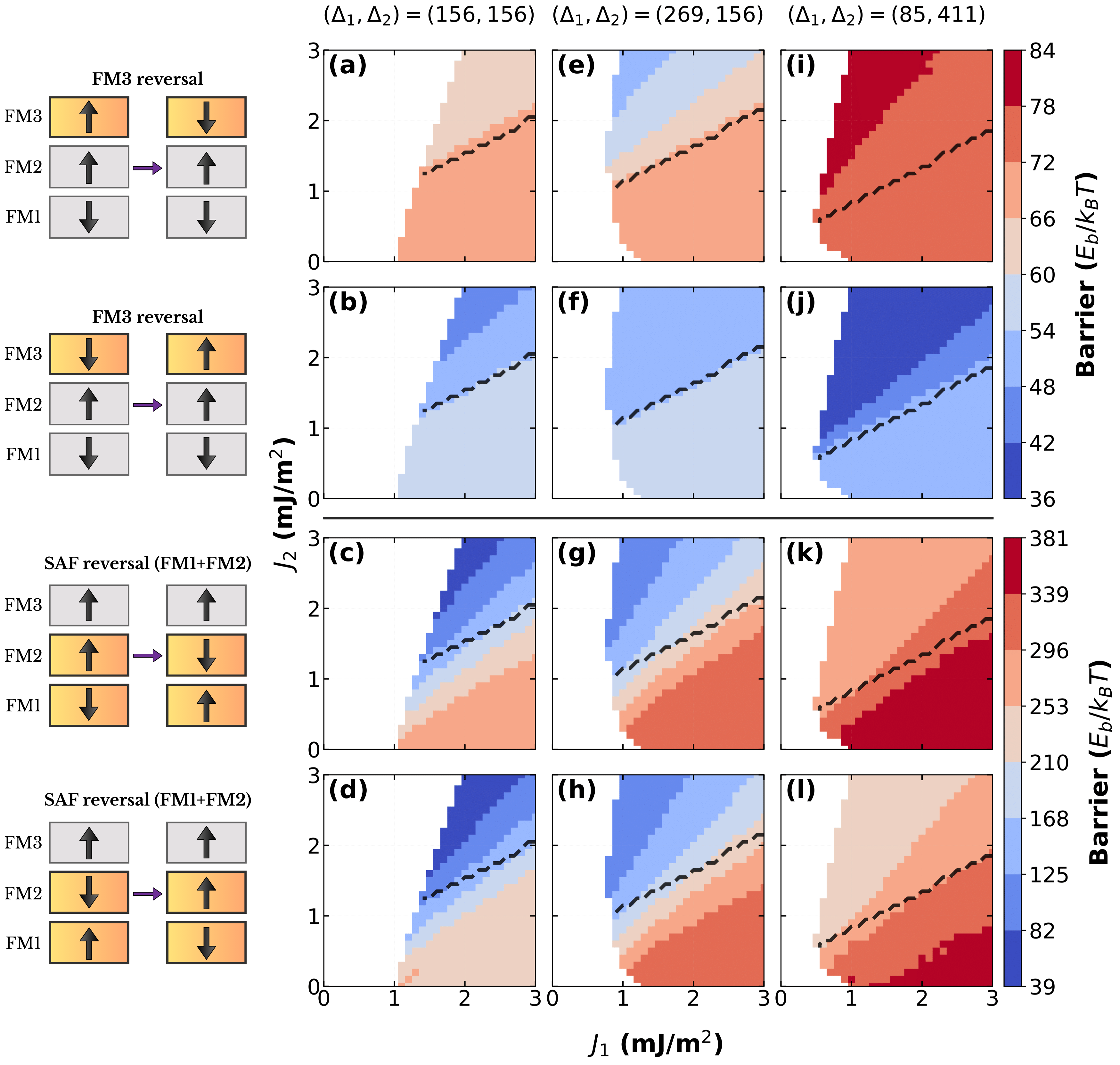}
    \caption{Switching-barrier maps as a function of interlayer exchange couplings $J_1$ and $J_2$ for three parameter sets, computed with the string method. FM3 parameters are identical in all three cases. Columns correspond to the parameter sets in \cref{fig:magnetic_config}(a--c): $(\Delta_1, \Delta_2) = (156, 156)$ for panels~\textbf{(a--d)}, $(269, 156)$ for panels~\textbf{(e--h)}, and $(85, 411)$ for panels~\textbf{(i--l)}. Rows~1--2 show the FM3 reversal barrier for two initial-state branches, distinguished by whether the net SAF stray field stabilizes or destabilizes the initial FM3 orientation. Rows~3--4 show the SAF reversal barrier for the same two branches. Barriers are reported as $E_{\mathrm{b}}/(k_{\mathrm{B}}T)$ at $T = 300~\mathrm{K}$, using micromagnetic energies without thermal fluctuations. Cartoons at left indicate the initial and final layer orientations (FM1 bottom, FM3 top), with highlighted layers indicating the switching subsystem. The dashed contour marks the boundary between APc and APnc. Separate color scales are used for the FM3-reversal and SAF-reversal rows.}
    \label{fig:side_by_side_barriers}
\end{figure*}

We now map the FM3 and SAF reversal barriers across the APc and APnc portions of the $(J_1, J_2)$ space for three representative parameter sets from \cref{fig:magnetic_config}(a--c): $(\Delta_1, \Delta_2) = (156, 156)$, $(269, 156)$, and $(85, 411)$ (\cref{tab:combined_material_parameters}). FM3 ($M_{s,3} = 1000~\mathrm{kA\,m^{-1}}$, $t_3 = 1.3~\mathrm{nm}$, $\Delta_3 = 60$) is the same in all three cases. For each $(J_1, J_2)$ point in these APc and APnc regions, we compute the two barrier heights (one for each direction) for both FM3 reversal and SAF reversal via the string method. We restrict the calculations to APc and APnc structures because they are the target operating regimes for p-STT-MRAM. The remaining phases in these three structures have 6 or 8 minima, and a systematic barrier analysis of these mixed-minimum regions is beyond the scope of this work. The resulting maps are shown in \cref{fig:side_by_side_barriers}.

\emph{FM3 reversal barriers.} The same ordering $E_{\mathrm{b}}^{\downarrow\uparrow\uparrow \to \downarrow\uparrow\downarrow} > 60\,k_{\mathrm{B}}T > E_{\mathrm{b}}^{\downarrow\uparrow\downarrow \to \downarrow\uparrow\uparrow}$ holds in the APc region for all three $(\Delta_1, \Delta_2)$ parameter sets. In APc configurations, FM1 and FM2 remain nearly antiparallel, so the net stray field on FM3 is nearly independent of $(J_1, J_2)$ and the FM3 barriers are essentially constant for each parameter set. A small residual variation appears close to the APc/APnc boundary because our APc/APnc classification allows a finite collinearity tolerance (\cref{subsec:class}). The $(156, 156)$ and $(269, 156)$ parameter sets share the same $M_s$ and $t$ for both SAF layers and differ only in $K_u$, so the stray field and FM3 barriers in APc are effectively the same in both cases (\cref{fig:side_by_side_barriers}(a, b, e, f)). For the $(85, 411)$ set, however, FM2 has a larger $M_s t$ product and is closer to FM3 (\cref{tab:combined_material_parameters}), resulting in a larger SAF stray field on FM3. This widens the spread in barrier energies: $E_{\mathrm{b}}^{\downarrow\uparrow\uparrow \to \downarrow\uparrow\downarrow}$ increases and $E_{\mathrm{b}}^{\downarrow\uparrow\downarrow \to \downarrow\uparrow\uparrow}$ decreases (\cref{fig:side_by_side_barriers}(i, j)) relative to the other two sets. Specifically, $E_{\mathrm{b}}^{\downarrow\uparrow\uparrow \to \downarrow\uparrow\downarrow}$ is $71$, $71$, and $77\,k_{\mathrm{B}}T$ for $(156, 156)$, $(269, 156)$, and $(85, 411)$, respectively, and $E_{\mathrm{b}}^{\downarrow\uparrow\downarrow \to \downarrow\uparrow\uparrow}$ is $57$, $57$, and $52\,k_{\mathrm{B}}T$.

In the APnc region, the equilibrium angle between FM1 and FM2 depends on $(J_1, J_2)$. The resulting FM3 barrier trends reflect the combined effects of $J_1$, $J_2$, $K_u$, $M_s$, layer thicknesses, spacings, and positions that set the dipolar field. Higher $K_u$ generally correlates with smaller tilt and a larger retained magnetization component along $\hat{\mathbf{z}}$, but this trend is modulated by the other parameters. The FM3 barriers therefore vary with $(J_1, J_2)$. For $(\Delta_1, \Delta_2) = (156, 156)$, both SAF layers have the same $K_u$, so their tilts are nearly equal and both FM3 barriers decrease with increasing $J_2$ (\cref{fig:side_by_side_barriers}(a, b)). For $(269, 156)$, FM1 has the higher $K_u$ and tilts less. Both barriers still decrease with $J_2$ (\cref{fig:side_by_side_barriers}(e, f)), even though the magnetization of FM2 tilts more than that of FM1 with increasing $J_2$. For $(\Delta_1, \Delta_2) = (85, 411)$, FM1 and FM2 differ in all three parameters ($M_s$, $t$, and $K_u$). Because FM2 has the higher $K_u$ and is thicker, its magnetization tilts less than that of FM1 as $J_2$ increases, while FM1 tilts more and cancels less of the FM2 stray field. This results in an increase of the net stray field on FM3 with increasing $J_2$, raising $E_{\mathrm{b}}^{\downarrow\uparrow\uparrow \to \downarrow\uparrow\downarrow}$ while lowering $E_{\mathrm{b}}^{\downarrow\uparrow\downarrow \to \downarrow\uparrow\uparrow}$. This parameter set produces the widest spread of barriers across $(J_1, J_2)$ and sets the extremes of the FM3 barrier color scale ($36$ to $84\,k_{\mathrm{B}}T$) at high $J_2$ (\cref{fig:side_by_side_barriers}(i, j)). The median of $E_{\mathrm{b}}^{\downarrow\uparrow\uparrow \to \downarrow\uparrow\downarrow}$ over all $(J_1, J_2)$ points in the APnc region is ${\sim}63$, ${\sim}59$, and ${\sim}77\,k_{\mathrm{B}}T$ for $(156, 156)$, $(269, 156)$, and $(85, 411)$, respectively, and that of $E_{\mathrm{b}}^{\downarrow\uparrow\downarrow \to \downarrow\uparrow\uparrow}$ is ${\sim}48$, ${\sim}49$, and ${\sim}40\,k_{\mathrm{B}}T$.

\emph{SAF reversal barriers.} Unlike the FM3 barriers, SAF barriers vary with the interlayer exchange couplings $(J_1, J_2)$ even in the APc region. Increasing $J_1$ raises the SAF barrier~\cite{10.1063/1.1467977} and increasing $J_2$ lowers it. The stray field from FM3 produces a similar but proportionally smaller effect on the SAF barrier, giving $E_{\mathrm{b}}^{\downarrow\uparrow\uparrow \to \uparrow\downarrow\uparrow} > E_{\mathrm{b}}^{\downarrow\uparrow\downarrow \to \uparrow\downarrow\downarrow}$ (\cref{fig:side_by_side_barriers}(c, g, k) and (d, h, l), respectively). For $(\Delta_1, \Delta_2) = (156, 156)$ at high $J_2$ in the APnc region, the SAF barrier can fall below the FM3 barrier, indicating that the SAF becomes less stable than FM3. For each parameter set, the lower SAF barrier is $E_{\mathrm{b}}^{\downarrow\uparrow\downarrow \to \uparrow\downarrow\downarrow}$, which therefore limits SAF stability. Its median over all $(J_1, J_2)$ points in the APc region rises from ${\sim}242\,k_{\mathrm{B}}T$ for $(156, 156)$ to ${\sim}299\,k_{\mathrm{B}}T$ for $(269, 156)$ and ${\sim}331\,k_{\mathrm{B}}T$ for $(85, 411)$, and in the APnc region from ${\sim}90\,k_{\mathrm{B}}T$ to ${\sim}144\,k_{\mathrm{B}}T$ and ${\sim}248\,k_{\mathrm{B}}T$, respectively.

The APnc regions show a consistent pattern: the lower SAF barrier ($E_{\mathrm{b}}^{\downarrow\uparrow\downarrow \to \uparrow\downarrow\downarrow}$) increases across the three sets, while the lower FM3 barrier ($E_{\mathrm{b}}^{\downarrow\uparrow\downarrow \to \downarrow\uparrow\uparrow}$) has a median below $50\,k_{\mathrm{B}}T$ in all three sets. Additionally, improving SAF stability in APnc comes at the cost of reduced FM3 stability. In APnc the separation between the two FM3 barriers is smaller for $(156, 156)$ and $(269, 156)$, and larger for $(85, 411)$, where the higher-$K_u$, thicker layer (FM2) is adjacent to FM3 and exerts a stronger stray field. This trade-off is not observed in APc, where the FM3 barriers are nearly constant and the SAF barriers remain well above the FM3 values. The reduced FM3 stability in APnc could be offset by adjusting the FM3 layer properties ($M_{s,3}$, $t_3$, $K_{u,3}$) or by tuning the FM1/FM2 spacer and FM2/FM3 tunnel-barrier thicknesses, both of which were held fixed in this work.

\section{Conclusion}

We used finite-element micromagnetic simulations to map static equilibrium configurations of $30~\mathrm{nm}$-diameter FM1/FM2/FM3 p-STT-MRAM nanopillars as functions of the bilinear and biquadratic interlayer exchange coupling coefficients, $(J_1,J_2)$, between FM1 and FM2. The simulations also varied the $M_s$, $K_u$, and thickness of FM1, FM2, and FM3. We identified APc, APnc, Pc, and Pnc configuration groups as well as regions where multiple groups coexist.

The configurations of interest for p-STT-MRAM are APc-only and APnc-only regions, in which all relaxed minima have antiparallel FM1 and FM2 out-of-plane components, thereby minimizing the stray field acting on FM3. In these regions, Pc and Pnc minima are absent from the simulated energy landscape, removing competing parallel-SAF configurations that can produce larger uncompensated stray fields on FM3. Within the material and coupling ranges studied here, introducing asymmetry between FM1 and FM2 through $M_s$, $K_u$, or thickness shifts the APc-only and APnc-only regions toward lower coupling strengths. This trend is useful for stacks in which the available antiferromagnetic IEC strength is limited, for example after high-temperature annealing of magnetic tunnel junctions~\cite{tomczak_influence_2016,lee_effect_2016,liu_strong_2019,nakano_annealing_2018}.

The effect of SAF asymmetry is different for APc and APnc states. APnc-only regions expand strongly when the FM1 and FM2 parameters are asymmetric, whereas APc-only regions are less sensitive to asymmetry. When the analysis is restricted to weak bilinear coupling, APc-only regions appear in the sampled parameter set only when at least one SAF layer has a comparatively low stability factor. The stability factors $(\Delta_1,\Delta_2)$ provide a compact way to organize these trends since they depend on the key material parameters $M_s$, $K_u$, and layer thickness that govern the static magnetic configuration of p-STT-MRAM. The free layer also modifies the SAF phase diagram through its stray field. In the two FM3 cases compared here, the thicker, higher-$M_s$ free layer reduces the size of APc-only and APnc-only regions, with stronger suppression for APnc than for APc.

Minimum-energy path calculations in the APc-only and APnc-only regions show that the SAF and FM3 reversal barriers are coupled through dipolar fields. SAF stray fields split the two FM3 reversal barriers depending on whether the net dipolar field assists or opposes FM3 reversal, while the stray field from FM3 similarly splits the two SAF reversal barriers. In APc regions, the FM3 barriers are nearly independent of $(J_1,J_2)$ for each parameter set, although they are shifted from the isolated-FM3 value by the SAF field. In APnc regions, the FM1--FM2 angle changes across the phase diagram, and both SAF and FM3 barriers vary more strongly with coupling.

The barrier maps identify a design trade-off for APnc structures. Increasing SAF asymmetry can improve the stability of APnc reference states, but it can also reduce one branch of the FM3 reversal barrier. This trade-off is much weaker in the APc cases studied here, where the FM3 barriers are more predictable across the APc region and the SAF barriers remain well above the FM3 barriers. APc alignment is therefore the simpler design regime when the main goal is robust stability of both the antiparallel reference layer and the free layer.

APnc alignment should nevertheless be viewed as a useful design option rather than only as a competing state. Its built-in noncollinearity between FM2 and FM3 provides a finite initial spin-transfer torque, a mechanism associated with lower switching currents and shorter switching times~\cite{Sbiaa_2013}. Future current-driven simulations can quantify this switching behavior while incorporating the SAF and FM3 energy-barrier trends identified here.

Overall, these results show that SAF alignment is coupled to both the SAF and FM3 reversal barriers in the nanopillar stack. The dataset released with this work provides equilibrium states, energies, and layer-resolved magnetization angles across the simulated configurations, supporting the design of p-STT-MRAM devices with either APc or APnc SAF reference states~\cite{terko_magnetization_dataset_2026}.

\section{Data Availability}
The complete dataset generated in this study, including the equilibrium magnetization states, energies, and layer-resolved magnetization angles for all simulations, is available from Zenodo~\cite{terko_magnetization_dataset_2026}. A public-access website is also provided for interactive viewing and plotting of these data~\cite{terko_sfu_2026}.

\begin{acknowledgments}
This research was enabled in part by support provided by Calcul Québec and the Digital Research Alliance of Canada. This research was funded in whole or in part by the Austrian Science Fund (FWF) Grants No. P 34671 and No. I 6068. We acknowledge the support of the Natural Sciences and Engineering Research Council of Canada (NSERC), funding reference number RGPIN-2025-06827. A.T. and G.L.-L. contributed equally to this work.
\end{acknowledgments}

\bibliography{references}

\end{document}